\documentclass[aps,prl,twocolumn,superscriptaddress,showpacs,nobalancelastpage]{revtex4}

\usepackage{graphicx}
\begin{document}


\title{Diffractive $\phi$-meson photoproduction on the proton near threshold}

\author{T.~Mibe}
  \altaffiliation[Present address: ]{Department of Physics and Astronomy, Ohio University, Athens, Ohio 45701}
  \affiliation{Research Center for Nuclear Physics, Osaka University, Ibaraki, Osaka 567-0047, Japan}
  \affiliation{Advanced Science Research Center, Japan Atomic Energy Research Institute, Tokai, Ibaraki 319-1195, Japan}
\author{W.C.~Chang}
  \affiliation{Institute of Physics, Academia Sinica, Taipei 11529, Taiwan}
\author{T.~Nakano}
  \affiliation{Research Center for Nuclear Physics, Osaka University, Ibaraki, Osaka 567-0047, Japan}
\author{D.S.~Ahn}
  \affiliation{Research Center for Nuclear Physics, Osaka University, Ibaraki, Osaka 567-0047, Japan}
  \affiliation{Department of Physics, Pusan National University, Busan 609-735, Korea}
\author{J.K.~Ahn}
  \affiliation{Department of Physics, Pusan National University, Busan 609-735, Korea}
\author{H.~Akimune}
  \affiliation{Department of Physics, Konan University, Kobe, Hyogo 658-8501, Japan}
\author{Y.~Asano}
  \affiliation{Synchrotron Radiation Research Center, Japan Atomic Energy Research Institute, Mikazuki, Hyogo 679-5198, Japan}
\author{S.~Date\'}
  \affiliation{Japan Synchrotron Radiation Research Institute, Mikazuki, Hyogo 679-5198, Japan}
\author{H.~Ejiri}
  \affiliation{Research Center for Nuclear Physics, Osaka University, Ibaraki, Osaka 567-0047, Japan}
  \affiliation{Japan Synchrotron Radiation Research Institute, Mikazuki, Hyogo 679-5198, Japan}
\author{H.~Fujimura}
  \altaffiliation[Present address: ]{Department of Physics, Kyoto University, Kyoto 606-8502, Japan} 
  \affiliation{School of Physics, Seoul National University, Seoul, 151-747, Korea}
\author{M.~Fujiwara}
  \affiliation{Research Center for Nuclear Physics, Osaka University, Ibaraki, Osaka 567-0047, Japan}
  \affiliation{Advanced Photon Research Center, Japan Atomic Energy Research Institute, Kizu, Kyoto, 619-0215, Japan}
\author{K.~Hicks}
  \affiliation{Department of Physics and Astronomy, Ohio University, Athens, Ohio 45701, USA}
\author{T.~Hotta}
  \affiliation{Research Center for Nuclear Physics, Osaka University, Ibaraki, Osaka 567-0047, Japan}
\author{K.~Imai}
  \affiliation{Department of Physics, Kyoto University, Kyoto 606-8502, Japan} 
\author{T.~Ishikawa}
  \altaffiliation[Present address: ]{Laboratory of Nuclear Science, Tohoku University, Sendai, Miyagi 982-0826, Japan}
  \affiliation{Department of Physics, Kyoto University, Kyoto 606-8502, Japan} 
\author{T.~Iwata}
  \affiliation{Department of Physics, Yamagata University, Yamagata 990-8560, Japan}
\author{H.~Kawai}
  \affiliation{Department of Physics, Chiba University, Chiba 263-8522, Japan}
\author{Z.Y.~Kim}
  \affiliation{School of Physics, Seoul National University, Seoul, 151-747, Korea}
\author{K.~Kino}
  \altaffiliation[Present address: ]{Center for Nuclear Study, University of Tokyo,7-3-1 Hongo, Bunkyo, Tokyo 113-0033, Japan}
  \affiliation{Research Center for Nuclear Physics, Osaka University, Ibaraki, Osaka 567-0047, Japan}
\author{H.~Kohri}
  \affiliation{Research Center for Nuclear Physics, Osaka University, Ibaraki, Osaka 567-0047, Japan}
\author{N.~Kumagai}
  \affiliation{Japan Synchrotron Radiation Research Institute, Mikazuki, Hyogo 679-5198, Japan}
\author{S.~Makino}
  \affiliation{Wakayama Medical University, Wakayama, 641-8509, Japan}
\author{T.~Matsuda}
  \affiliation{Department of Applied Physics, Miyazaki University, Miyazaki 889-2192, Japan}
\author{T.~Matsumura}
  \altaffiliation[Present address: ]{Department of Applied Physics, National Defense Academy, Yokosuka 239-8686, Japan}
  \affiliation{Research Center for Nuclear Physics, Osaka University, Ibaraki, Osaka 567-0047, Japan}
  \affiliation{Advanced Science Research Center, Japan Atomic Energy Research Institute, Tokai, Ibaraki 319-1195, Japan}
\author{N.~Matsuoka}
  \affiliation{Research Center for Nuclear Physics, Osaka University, Ibaraki, Osaka 567-0047, Japan}
\author{K.~Miwa}
  \affiliation{Department of Physics, Kyoto University, Kyoto 606-8502, Japan} 
\author{M.~Miyabe}
  \affiliation{Department of Physics, Kyoto University, Kyoto 606-8502, Japan} 
\author{Y.~Miyachi}
  \altaffiliation[Present address: ]{Department of Physics, Tokyo Institute of Technology, Tokyo 152-8551, Japan}
  \affiliation{Department of Physics and Astrophysics, Nagoya University, Nagoya, Aichi 464-8602, Japan}
\author{M.~Morita}
  \affiliation{Research Center for Nuclear Physics, Osaka University, Ibaraki, Osaka 567-0047, Japan}
\author{N.~Muramatsu}
  \affiliation{Research Center for Nuclear Physics, Osaka University, Ibaraki, Osaka 567-0047, Japan}
\author{M.~Niiyama}
  \affiliation{Department of Physics, Kyoto University, Kyoto 606-8502, Japan} 
\author{M.~Nomachi}
  \affiliation{Department of Physics, Osaka University, Toyonaka, Osaka 560-0043, Japan}
\author{Y.~Ohashi}
  \affiliation{Japan Synchrotron Radiation Research Institute, Mikazuki, Hyogo 679-5198, Japan}
\author{T.~Ooba}
  \affiliation{Department of Physics, Chiba University, Chiba 263-8522, Japan}
\author{H.~Ohkuma}
  \affiliation{Japan Synchrotron Radiation Research Institute, Mikazuki, Hyogo 679-5198, Japan}
\author{D.S.~Oshuev}
  \affiliation{Institute of Physics, Academia Sinica, Taipei 11529, Taiwan}
\author{C.~Rangacharyulu}
  \affiliation{Department of Physics and Engineering Physics, University of Saskatchewan, Saskatoon, Saskatchewan, Canada, S7N 5E2} 
\author{A.~Sakaguchi}
  \affiliation{Department of Physics, Osaka University, Toyonaka, Osaka 560-0043, Japan}
\author{T.~Sasaki}
  \affiliation{Department of Physics, Kyoto University, Kyoto 606-8502, Japan} 
\author{P.M.~Shagin}
  \altaffiliation[Present address: ]{ Department of Physics and Astronomy, Rice University, 6100 Main St. Houston MS 108, TX 77005-1892, USA}
  \affiliation{Research Center for Nuclear Physics, Osaka University, Ibaraki, Osaka 567-0047, Japan}
\author{Y.~Shiino}
  \affiliation{Department of Physics, Chiba University, Chiba 263-8522, Japan}
\author{H.~Shimizu}
  \affiliation{Laboratory of Nuclear Science, Tohoku University, Sendai, Miyagi 982-0826, Japan}
\author{Y.~Sugaya}
  \affiliation{Department of Physics, Osaka University, Toyonaka, Osaka 560-0043, Japan}
  \affiliation{Advanced Science Research Center, Japan Atomic Energy Research Institute, Tokai, Ibaraki 319-1195, Japan}
\author{M.~Sumihama}
  \altaffiliation[Present address: ]{Department of Physics, Tohoku University, Sendai, Miyagi 980-8578, Japan}
  \affiliation{Department of Physics, Osaka University, Toyonaka, Osaka 560-0043, Japan}
  \affiliation{Advanced Science Research Center, Japan Atomic Energy Research Institute, Tokai, Ibaraki 319-1195, Japan}
\author{A.I.~Titov}
  \altaffiliation[Present address: ]{Joint Institute for Nuclear Research, 141980, Dubna, Russia}
  \affiliation{Advanced Photon Research Center, Japan Atomic Energy Research Institute, Kizu, Kyoto, 619-0215, Japan}
\author{Y.~Toi}
  \affiliation{Department of Applied Physics, Miyazaki University, Miyazaki 889-2192, Japan}
\author{H.~Toyokawa}
  \affiliation{Japan Synchrotron Radiation Research Institute, Mikazuki, Hyogo 679-5198, Japan}
\author{A.~Wakai}
  \altaffiliation[Present address: ]{Akita Research Institute of Brain and Blood Vessels, Akita 010-0874, Japan}
  \affiliation{Center for Integrated Research in Science and Engineering, Nagoya University, Nagoya, Aichi 464-8603, Japan}
\author{C.W.~Wang}
  \affiliation{Institute of Physics, Academia Sinica, Taipei 11529, Taiwan}
\author{S.C.~Wang}
  \altaffiliation[Present address: ]{Institute of Statistical Science, Academia Sinica, Nankang, 115 Taipei, Taiwan}
  \affiliation{Institute of Physics, Academia Sinica, Taipei 11529, Taiwan}
\author{K.~Yonehara}
  \altaffiliation[Present address: ]{Illinois Institute of Technology, Chicago, Illinois 60616, USA}
  \affiliation{Department of Physics, Konan University, Kobe, Hyogo 658-8501, Japan}
\author{T.~Yorita}
  \affiliation{Japan Synchrotron Radiation Research Institute, Mikazuki, Hyogo 679-5198, Japan}
\author{M.~Yoshimura}
  \affiliation{Institute for Protein Research, Osaka University, Suita, Osaka 565-0871, Japan}
\author{M.~Yosoi}
  \altaffiliation[Present address: ]{Research Center for Nuclear Physics, Osaka University, Ibaraki, Osaka 567-0047, Japan}
  \affiliation{Department of Physics, Kyoto University, Kyoto 606-8502, Japan} 
\author{R.G.T.~Zegers}
  \altaffiliation[Present address: ]{National Superconducting Cyclotron Laboratory, Michigan State University, East lansing, MI 48824-1321, USA}
  \affiliation{Research Center for Nuclear Physics, Osaka University, Ibaraki, Osaka 567-0047, Japan}

\collaboration{The LEPS collaboration}
\noaffiliation

\date{October 7, 2005}

\begin{abstract}
 Photoproduction of $\phi$-meson on protons was studied by means of linearly
polarized photons at forward angles in the low-energy region from 
threshold to $E_{\gamma}$= 2.37 GeV.  The differential cross sections at $t = -|t|_{min}$ do not
increase smoothly as $E_{\gamma}$ increases, but show a local maximum at around 2.0
GeV.  The angular distributions demonstrate that $\phi$-mesons are photo-produced
predominantly by helicity-conserving processes, and the local maximum is
not likely due to unnatural-parity processes. 
\end{abstract}


\maketitle
The gluonic aspect of quantum chromodynamics (QCD), especially glueballs, has been 
of wide interest in hadron physics. 
It has been suggested that there is a connection between the glueball 
Regge trajectory ($J^{PC}=2^{++},4^{++}...$) and the Pomeron trajectory~\cite{donnachie}.
The diffractive photoproduction of $\phi$-mesons has traditionally been used 
to study the Pomeron exchange process~\cite{Bauer:1978iq}.
This is because the baryon and meson exchange amplitudes in the s- 
and t-channels are suppressed by the Okubo-Zweig-Iizuka (OZI) rule. 
Photoproduction of $\phi$-meson is useful to study 
not only Pomeron exchange, but also other hadronic interactions mediated by multi-gluon exchanges
which are difficult to identify in the other hadronic reactions due to large contributions from
baryon and meson exchanges.

 The low-energy diffractive photoproduction of 
$\phi$-mesons was suggested~\cite{Nakano:1997qd,Titov:1999eu} to be sensitive 
to a daughter Pomeron trajectory associated 
with a glueball ($J^{PC}=0^{++}$)~\cite{Nakano:1997qd,Titov:1999eu,Kisslinger:1999jk,Llanes-Estrada:2000jw}.
Its contribution is expected to decrease rapidly with an increase of photon
energy, whereas the contribution from Pomeron exchange increases.
 This difference may lead to a
non-monotonic energy dependence of the forward-angle cross section near threshold ($E_{\gamma}=1.57$ GeV).
The existing cross section data in the low-energy region~\cite{Ballam:1973eq,Besch:1974rp,Behrend:1978ik,Barber:1982fj,Anciant:2000az,Barth:2003bq}
are still too poor to ascertain a possible signature of such a non-monotonic behavior.

The background contributions from s- and u-channels
diagrams, such as direct $\phi$ radiation from the nucleon~\cite{Titov:1999eu,Oh:2001bq}
and production in nucleon resonance decay~\cite{Zhao:2001ue,Titov:2003bk}, are 
predicted to be small at small $|t|$ $(t=(p_{\phi}-p_{\gamma})^2)$.
However, the contributions from the t-channel exchanges of pseudoscalar mesons ($\pi,\eta$),
scalar mesons ($f_0$, $a_0$)~\cite{Titov:1999eu,Williams:1998ge} and a tensor $f_2'$ meson~\cite{Laget:2000gj}
are predicted not to be negligible.
The energy dependence of those meson-exchange processes is expected to be similar to that of the daughter Pomeron trajectory.
Therefore, to determine the relative contributions of these processes, we need to
analyze the spin observables using linearly polarized photons.

The spin observables are studied via the decay angular distribution of the $\phi$-meson
in the $K^+K^-$ decay mode.
 The decay angular distribution $W(\cos\theta,\phi,\Phi)$ is a function of the spin density matrix
 elements~\cite{Schilling:1970um}, where 
$\theta$ and $\phi$ denote the polar and azimuthal angles of the $K^+$ in the $\phi$-meson rest frame.
The azimuthal angle of the photon polarization in the center-of-mass frame is denoted by $\Phi$.
The relative contribution of natural-parity exchange and
 unnatural-parity exchange is related to the density matrix element
 ${\bar{\rho}}^1_{1-1}(\equiv
 1/2({\rho}^1_{1-1}-\mbox{Im}{\rho}^2_{1-1}))$
 which is extracted from the one-dimensional distribution $W(\phi-\Phi)$~\cite{Titov:2003bk} through 
\begin{eqnarray}
W(\phi-\Phi)= \frac{1}{2\pi}(1+2P_{\gamma}\bar{\rho}^1_{1-1}\cos 2(\phi-\Phi)),
\end{eqnarray}
where $P_{\gamma}$ is the polarization of the photon beam.
The available data at $E_{\gamma}=2.8,4.7,9.3$ GeV~\cite{Ballam:1973eq} and 20-40 GeV~\cite{Atkinson:1984cs} 
support the dominance of the helicity-conserving natural-parity exchange
processes. However, there is no measurement of polarization observables near the threshold. 

The decay angular distributions also provide information on helicity non-conserving processes.
Recent measurements at low energies with unpolarized photons suggest 
significant contributions from the helicity non-conserving processes at large
momentum transfer $t$~\cite{{Barth:2003bq},{McCormick:2003bb}}.
The contribution from helicity non-conserving mechanisms is examined by a deviation from
the $\sin^2\theta$ behavior in the one dimensional distribution $W(\cos\theta)$
and an oscillation in the one dimensional distributions $W(\phi+\Phi), W(\phi)$
 and $W(\Phi)$.

In this Letter, we report measurements of the
differential cross sections ($d\sigma/dt$) at small $|t|$ and 
the first measurements of decay angular distributions near threshold with linearly
polarized photons.
Linearly polarized photons were produced 
by means of the backward-Compton scattering of laser photons off the 8 GeV electron
at the SPring-8 BL33LEP beamline (LEPS: Laser Electron Photons at SPring-8 facility)~\cite{Nakano:2001xp}.
The maximum energy of the photon beam was 2.4 GeV. 
The photon energy was determined by measuring
recoil electrons using a tagging counter with a resolution
($\sigma$) of 15 MeV. The typical photon flux was about $10^6$ s$^{-1}$,
which was monitored by counting scattered electrons 
with the tagging system.
The systematic uncertainty in the photon flux measurement was estimated to be 3\%.
The degree of linear polarization varied
with photon energy; it was 95 \% at the maximum
energy, 60 \% at 1.57 GeV.
A liquid hydrogen target with a length of 50 mm was used in the experiment.
A similar experiment with nuclear targets 
and the associated analysis have been reported in Ref.~\cite{Ahn:2004id}.

The momenta and the time-of-flight (TOF) of produced charged particles 
were measured with a magnetic spectrometer~\cite{Nakano:2001xp}.
The angular coverage of the spectrometer is about $\pm 0.4$ rad
and $\pm 0.2$ rad in the horizontal and vertical directions,
respectively. The momentum resolution ($\sigma$) for 1 GeV/c particles was 6 MeV/c. 
The TOF resolution ($\sigma$) was 150 psec for a typical flight path length of
4 m. The mass resolution ($\sigma$) was 30 MeV/c$^2$ for a 1 GeV/c$^2$ kaon.
Pions with momenta higher than 0.6 GeV/c and $e^+e^-$ pairs were 
rejected by using an aerogel Cherenkov counter
in the trigger level. An over-veto rate in the trigger was estimated to be less 
than 2.1\%.

The incident photon energy and the momenta of 
$K^+K^-$ tracks or $K^{\pm}p$ tracks 
were measured to identify the reaction $\gamma p \rightarrow K^+K^-p$ followed by
the $\phi \rightarrow K^+K^-$ decay. 
Based on the detected particles, we define two
types of event topology: $K^+K^-$-reconstructed events ($KK$ mode),
and  $K^{\pm}p$-reconstructed events ($Kp$ mode).

\begin{figure}[b] 
\includegraphics[width=8.5cm]{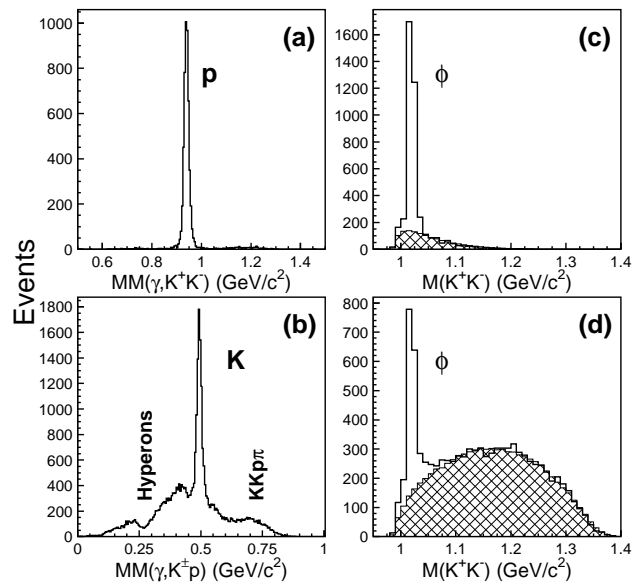} 
\caption{(a) Missing mass distribution for the $p(\gamma,K^+K^-)X$ reaction
 in $KK$ mode, (b) Missing mass distribution for
the $p(\gamma,K^{\pm}p)X$ reaction in $Kp$ mode.
(c) and (d) are the $K^+K^-$ invariant mass distributions after the cut on the missing mass 
for $KK$ and $Kp$ modes, respectively. The hatched histograms are the simulated background. 
} 
\label{fig.mass} 
\end{figure} 

\begin{figure*}[bth] 
\includegraphics[width=17.0cm]{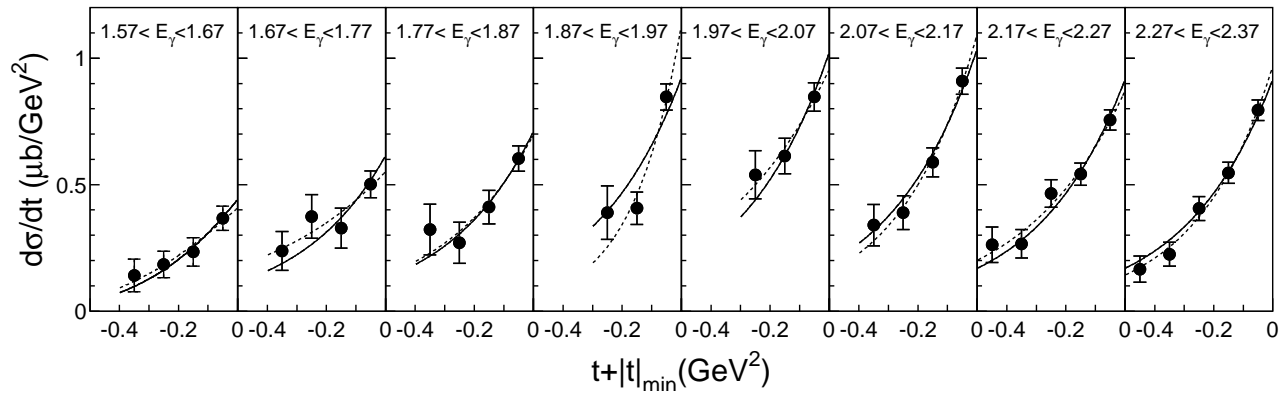} 
\caption{Differential cross sections for the $\gamma p \rightarrow \phi p$ reaction.
 The dashed curves are the results of the fit using an exponential function 
($(d\sigma/dt)_{t=-|t|_{min}}\mbox{e}^{b(t+|t|_{min})}$) with $(d\sigma/dt)_{t=-|t|_{min}}$ and $b$ as free parameters.
The solid curves are fitted results with fixing $b= 3.38$ GeV$^{-2}$.
The error bars represent statistical errors. The systematic errors are discussed in the text.
}
\label{fig.dndt} 
\end{figure*} 

 The missing mass distribution for the $p(\gamma,K^+K^-)X$ reaction 
(denoted as $MM(\gamma,K^+K^-)$) is shown in
Fig.~\ref{fig.mass}(a) for $KK$ mode. A sharp
peak corresponding to the proton was observed with an average mass resolution ($\sigma$) of 10 MeV/c$^2$.
The missing mass distribution for the $p(\gamma,K^{\pm}p)X$ reaction ($MM(\gamma,K^{\pm}p)$) is shown
in Fig.~\ref{fig.mass}(b) for $Kp$ mode. A clear peak at
the kaon mass was observed with an average mass resolution ($\sigma$) of 10 MeV/c$^2$.
 In $Kp$ mode, there were contributions from
 non-$K^+K^-p$ final states. The background below the kaon peak is
 attributed mainly to hyperon photoproduction having a non-$K^+K^-p$
 final state, such as $Kp\pi\gamma$ , $Kp\pi\pi$. The background above the kaon peak 
is due to $KKp\pi$ events. A 3~$\sigma$ cut on the missing mass spectrum was applied 
to select the $K^+K^-p$ final state.

Figure~\ref{fig.mass}(c) and (d) show the $K^+K^-$ invariant mass
distributions for $KK$ and $Kp$ modes, respectively. 
In $Kp$ mode, the momentum of the missing kaon was calculated 
by assuming a $K^+K^-p$ final state. 
The cut point on the $K^+K^-$ invariant mass was set to $1.009<M(K^+K^-)<1.029$
GeV/c$^2$ which corresponded to about 10~\% loss of $\phi$ events.
The background in the $\phi$ peak region was estimated with the following method.
We considered two sources of background; photoproduction 
of $\Lambda(1520)$ and  a $K^+K^-p$
 final state without forming any narrow resonance structure
 in either $K^+K^-$ or $K^{\pm}p$ system (non-resonant $KKp$). 
 The background level was estimated from the yields below and above the 
$\phi$-meson peak by using 
 Monte Carlo simulations which were fitted to the angular distributions of $K^+$, $K^-$ and $p$ in the real data. 
 The Monte Carlo simulations reproduced the $K^+K^-$ invariant mass (Fig.~\ref{fig.mass}(c) and (d)) 
 and $K^-p$ invariant mass distributions in the real data.
 Although there is a kinematical overlap of $\Lambda(1520)$ and $\phi$ production in this energy range,
 the contamination of $\Lambda(1520)$ events in the final sample is suppressed in small $|t|$ regions.
 It was estimated as less than a few percent.
 The estimated systematic error on the cross section due to
 the background subtraction procedure was less than 0.8\%.

The acceptance of the spectrometer was determined in Monte Carlo simulations
using the GEANT3 simulation package~\cite{Brun:1978fy}. 
Geometrical acceptance, resolution and
efficiency of the detectors were taken into account. Since the acceptance
depends on the input distributions, the simulations were iterated to
reflect measured $d\sigma/dt$ and angular distributions, having started from
flat distributions. The acceptance also depends on beam polarization and the
mode of reconstruction. 
 The validity of the acceptance calculation and the background
 subtraction was confirmed by checking the 
 consistency of the cross section results among different reconstruction
 modes, and also by checking the consistency of the decay angular distributions 
 obtained with different beam polarizations. 

The differential cross sections were measured in terms of 
$t+|t|_{min}$ where $|t|_{min}$ is the minimum 4-momentum transfer from the incident photon to the $\phi$-meson.
Figure~\ref{fig.dndt} shows the $d\sigma/dt$ in
different photon energy regions. 
The $d\sigma/dt$ showed a forward peaking shape,
suggesting the dominance of t-channel exchange processes.
A fit to the $d\sigma/dt$ was performed
with an exponential function; i.e. $(d\sigma/dt)_{t=-|t|_{min}}\mbox{e}^{b(t+|t|_{min})}$
with $(d\sigma/dt)_{t=-|t|_{min}}$ and $b$ as free
 parameters. 
No strong energy dependence of the slope $b$ was found beyond statistical errors.
The average value of the slope $b$ was $3.38\pm 0.23$ GeV$^{-2}$.
When the average slope for data at all energies was used in the fit, 
the fitting curves described the data points well.

\begin{figure}[tb] 
\includegraphics[width=8.5cm]{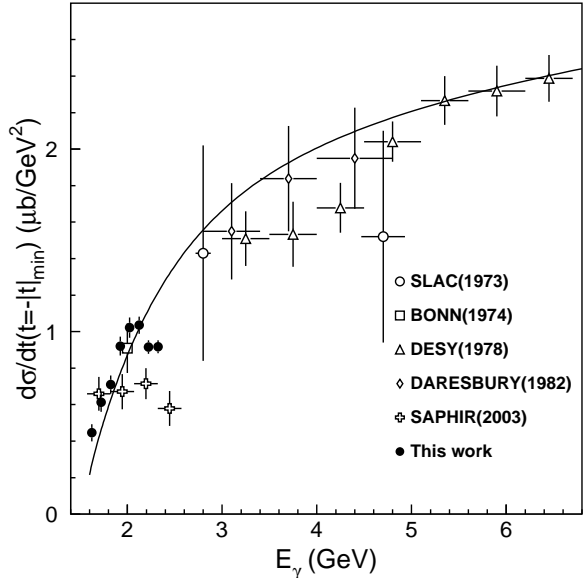}
\caption{Energy dependence of $(d\sigma/dt)_{t=-|t|_{min}}$.
The closed circles are the results of the present work. Other data points are taken from Ref.~\cite{Ballam:1973eq,Besch:1974rp,Behrend:1978ik,Barber:1982fj,Anciant:2000az,Barth:2003bq}.
The error bars represent statistical errors. The systematic errors are discussed in the text.
The solid curve represents the prediction of a model including the Pomeron trajectory,
$\pi$ and $\eta$ exchange processes~\cite{Titov:2003bk}.
}

\label{fig.crosseg} 
\end{figure}

Figure~\ref{fig.crosseg} shows the energy dependence of 
$(d\sigma/dt)_{t=-|t|_{min}}$ when $b$ is set to the average slope.
The energy dependence of $(d\sigma/dt)_{t=-|t|_{min}}$
shows a non-monotonic behavior with a
local maximum at $E_{\gamma} \sim 2$ GeV. 
The local maximum is also seen using only 
the differential cross sections at the lowest $t+|t|_{min}$ bin 
where acceptance and signal-to-noise ratio is at a maximum. The local maximum still persisted when
the analysis was repeated excluding events near the $\Lambda(1520)$ peak in $K^-p$ system.

Data were compared with the prediction of a model including Pomeron exchange,
$\pi$ and $\eta$ exchange processes~\cite{Titov:2003bk}.
A $\chi^2$ test was performed to check whether the model prediction was statistically compatible.
It gave $\chi^2= 140$ for 8 degrees of freedom using the present measurements.
The model is inconsistent with the present data points
although it describes the data rather well at higher energies. 

\begin{figure}[tb]
\includegraphics[width=8.5cm]{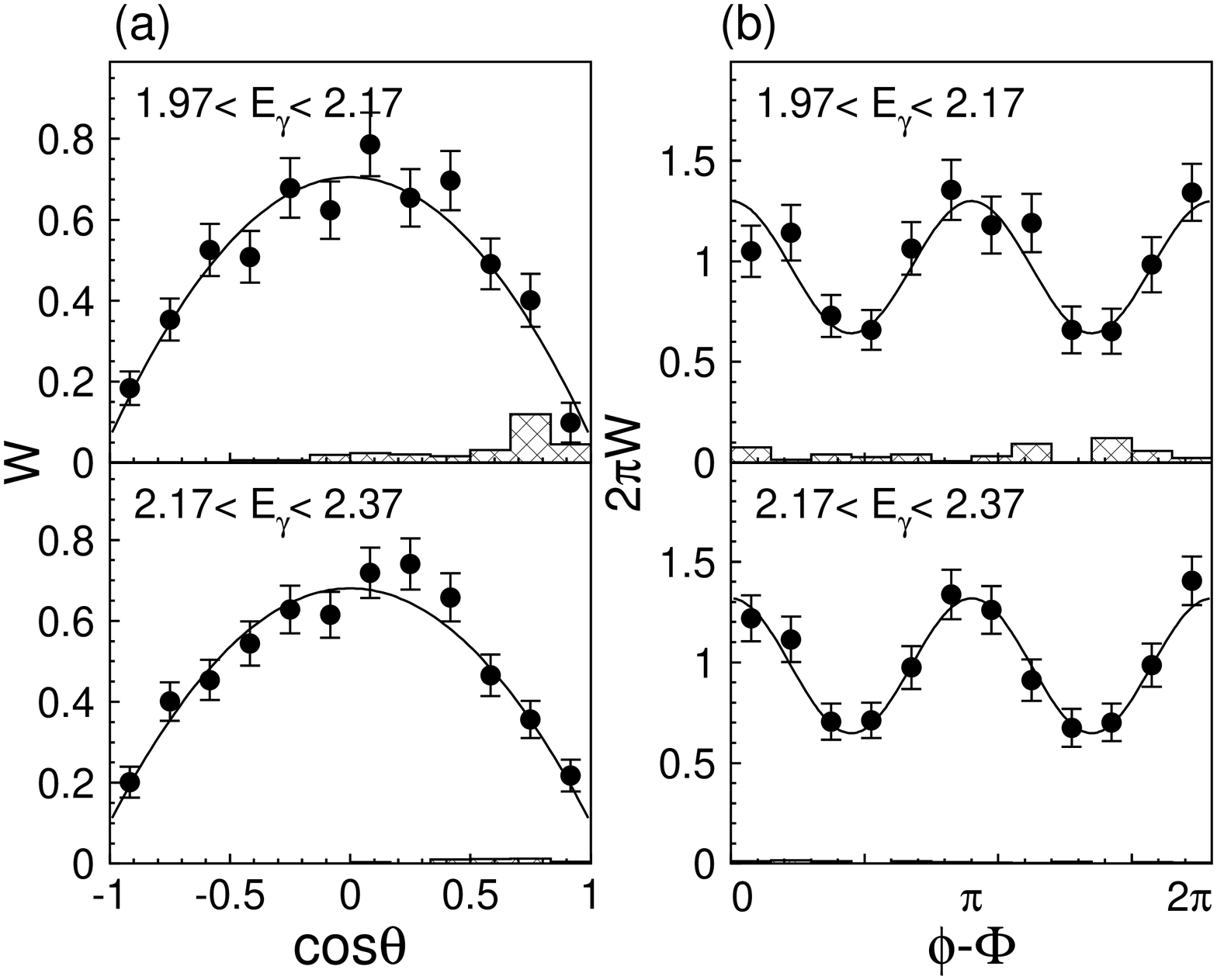}
\includegraphics[width=8.5cm]{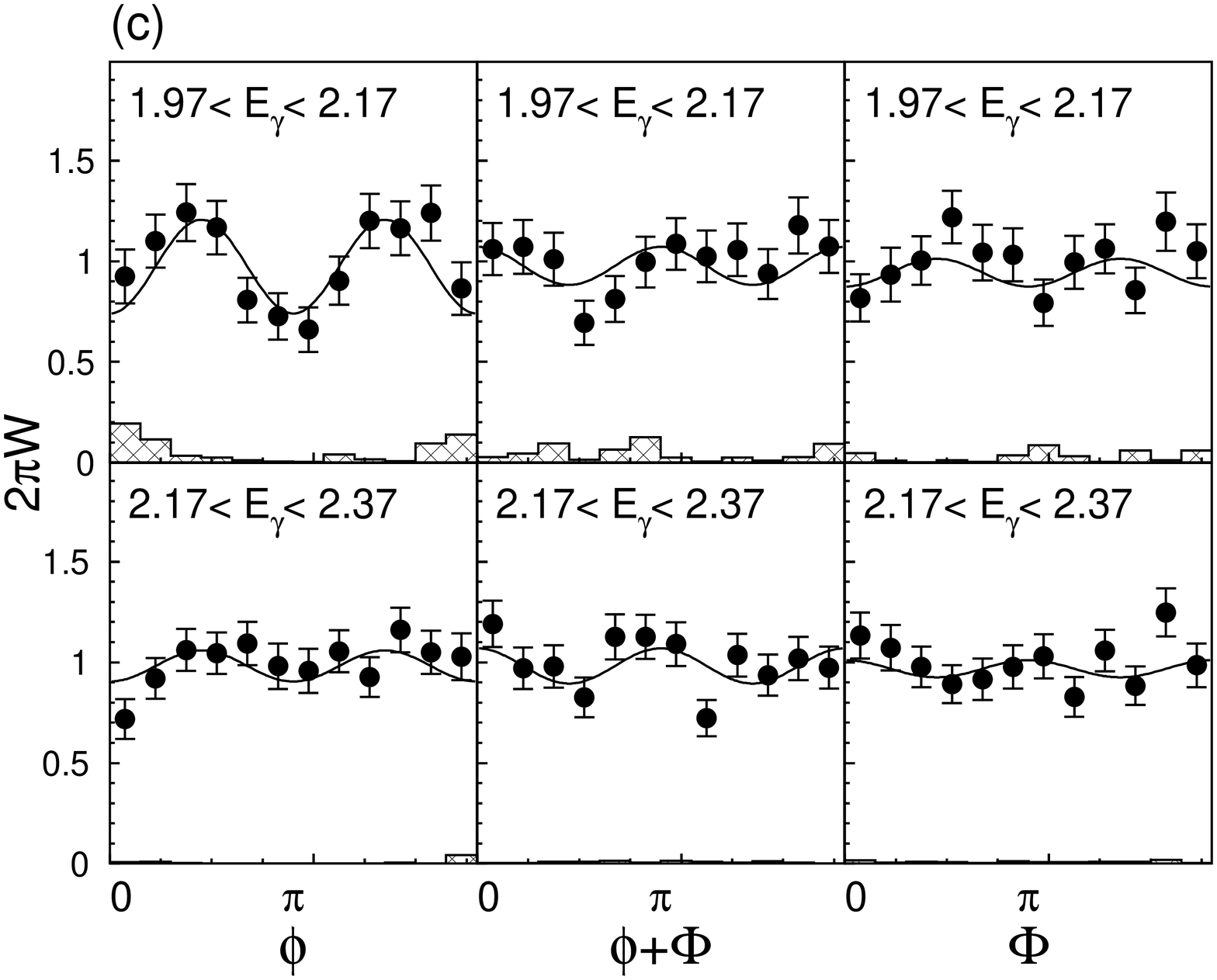}
\caption{Decay angular distributions for -0.2$<t+|t|_{min}$ in the Gottfried-Jackson frame. 
The solid curves are the fit to the data. The hatched histograms are systematic errors.}
\label{fig.angleavflip}
\end{figure}

 The decay angular distributions in the Gottfried-Jackson frame 
were obtained at forward angles
 ($-0.2<t+|t|_{min}\leq 0$ GeV$^2$) in two different energy-regions: (1)
 around the local maximum of the cross section ($\Delta E_1$:$1.97<E_{\gamma}<2.17$ GeV) and (2)
 above the local maximum ($\Delta E_2$:$2.17<E_{\gamma}<2.37$ GeV) where 
there are enough statistics and the acceptance is fairly flat over all angular variables.

 Figure~\ref{fig.angleavflip}(a) shows the angular
 distribution $W(\cos\theta)$.
 In both energy regions, $W(\cos\theta)$ behaves as $\sim(3/4)\,\sin^2\theta$,
 indicating the dominance of helicity-conserving processes. 
A contribution from tensor-meson exchange, such as $f_2'$-meson exchange,  must be small
since a contribution of this term would result in a deviation from the
$\sin^2\theta$ form~\cite{Titov:2003bk}. 
Note that this result is different from the measurement in a wider $t$ range~\cite{Barth:2003bq} which
shows strong violation from $\sin^2\theta$ form. This may be understood by the 
production mechanism discussed in Ref.~\cite{McCormick:2003bb}.

 Figure~\ref{fig.angleavflip}(b) shows the distribution $W(\phi-\Phi)$. We found
 ${\bar{\rho}}^1_{1-1}= 0.197\pm 0.030(\mbox{stat.})\pm 0.022(\mbox{sys.})$ in $\Delta E_1$
 and $0.189\pm 0.024(\mbox{stat.})\pm 0.006(\mbox{sys.})$ in $\Delta E_2$.
 The positive value for ${\bar{\rho}}^1_{1-1}$ indicates that the contributions
 from natural parity exchange are bigger than those for unnatural parity exchange ($\pi,\eta$-meson exchange).
 The ${\bar{\rho}}^1_{1-1}$ is the same in the two energy regions within errors.
 This implies that the relative contribution of natural parity
 exchange and unnatural parity exchange remains constant in the
 two energy regions. Therefore, it is difficult to attribute the origin of 
the local maximum in the cross
 section to different strengths of the unnatural-parity exchange processes in the two energy regions.

 Other one dimensional angular distributions $W(\phi)$, $W(\phi+\Phi)$ and $W(\Phi)$
 are depicted in Fig.~\ref{fig.angleavflip}(c). No strong oscillation was
 found, except that the distribution $W(\phi)$
 at $\Delta E_1$ bin showed an oscillation
 ($\rho^0_{1-1}=0.120\pm 0.027(\mbox{stat.})\pm 0.011(\mbox{sys.})$).
 $\rho^0_{1-1}$ reflects
 the double spin-flip transition from the incident photon to the outgoing $\phi$-meson~\cite{Schilling:1970um}.
 The spin-flip amplitudes are exactly zero in the case of pure scalar meson exchange and pseudoscalar meson exchange processes.
 The oscillation in the $W(\phi)$ distribution might be
 understood in the framework of a modified Donnachie-Landshoff Pomeron model
 motivated by the non-perturbative two-gluon-exchange dynamics~\cite{Titov:2003bk}.
 However, this model fails to reproduce the non-monotonic energy dependence
 (see the solid curve in Fig.~\ref{fig.crosseg}).

 An alternative explanation might be the manifestation of a daughter Pomeron 
trajectory. In this case,
 the decay angular distributions may be similar to those 
for the Pomeron trajectory as observed, since contributions 
from both trajectories involve exchanges of natural-parity particles.
 While the decay angular distributions are useful to discriminate natural parity exchange from
unnatural parity exchange, they are not useful to disentangle the two possible natural parity exchanges,
i.e. Pomeron exchange and a daughter Pomeron exchange. On the other hand, the energy dependence of the cross sections 
is a good indicator for a daughter Pomeron exchange process.  However, the fit suggested in
 Ref.~\cite{Nakano:1997qd} failed to predict the local maximum in the cross section
 with the proposed set of parameters. 

In summary, the photoproduction of the $\phi $-meson was studied for the
first time by means of linearly polarized photons at forward angles in the low energy
region from the threshold energy of $E_\gamma $=1.57 GeV to 2.37 GeV.  The
differential cross sections at $t=-|t|_{min}$ go 
non-monotonically as a function of $E_{\gamma}$, and show a local maximum at around 2.0 GeV.
The polar angle distributions demonstrate
dominance of helicity conserving processes and disfavor 
tensor $f'_2$ meson exchange. The azimuthal angle distributions
over the local maximum suggest that the local maximum is not due to additional
unnatural-parity processes, but likely due to new dynamics
which may involve a multi-gluon exchange 
beyond the Pomeron exchange process. 
Further theoretical and experimental studies are of great
interest for clarifying photoproduction mechanisms in the low-energy region
with the local maximum.

 The authors thank the SPring-8 staff
 for supporting the BL33LEP beam line and the LEPS experiment. We thank
 H. Toki and A. Hosaka (RCNP) for fruitful discussions. This
 research was supported in part by the Ministry of Education, Science,
 Sports and Culture of Japan, by the National Science Council of
 Republic of China (Taiwan), Korea Research Foundation(KRF) Grant(2003-015-C00130)
and National Science Foundation (NSF Award PHY-0244999).


\begin{thebibliography}{23}
\expandafter\ifx\csname natexlab\endcsname\relax\def\natexlab#1{#1}\fi
\expandafter\ifx\csname bibnamefont\endcsname\relax
  \def\bibnamefont#1{#1}\fi
\expandafter\ifx\csname bibfnamefont\endcsname\relax
  \def\bibfnamefont#1{#1}\fi
\expandafter\ifx\csname citenamefont\endcsname\relax
  \def\citenamefont#1{#1}\fi
\expandafter\ifx\csname url\endcsname\relax
  \def\url#1{\texttt{#1}}\fi
\expandafter\ifx\csname urlprefix\endcsname\relax\def\urlprefix{URL }\fi
\providecommand{\bibinfo}[2]{#2}
\providecommand{\eprint}[2][]{\url{#2}}

\bibitem[{\citenamefont{Donnachie et~al.}(2002)\citenamefont{Donnachie, Dosch,
  Landshoff, and Nachtmann}}]{donnachie}
\bibinfo{author}{\bibfnamefont{S.}~\bibnamefont{Donnachie}},
  \bibinfo{author}{\bibfnamefont{G.}~\bibnamefont{Dosch}},
  \bibinfo{author}{\bibfnamefont{P.}~\bibnamefont{Landshoff}},
  \bibnamefont{and}
  \bibinfo{author}{\bibfnamefont{O.}~\bibnamefont{Nachtmann}},
  \emph{\bibinfo{title}{Pomeron Physics and QCD}}
  (\bibinfo{publisher}{Cambridge University Press}, \bibinfo{year}{2002}),
  \bibinfo{note}{reference therein}.

\bibitem[{\citenamefont{Bauer et~al.}(1978)\citenamefont{Bauer, Spital, Yennie,
  and Pipkin}}]{Bauer:1978iq}
\bibinfo{author}{\bibfnamefont{T.~H.} \bibnamefont{Bauer}},
  \bibinfo{author}{\bibfnamefont{R.~D.} \bibnamefont{Spital}},
  \bibinfo{author}{\bibfnamefont{D.~R.} \bibnamefont{Yennie}},
  \bibnamefont{and} \bibinfo{author}{\bibfnamefont{F.~M.}
  \bibnamefont{Pipkin}}, \bibinfo{journal}{Rev. Mod. Phys.}
  \textbf{\bibinfo{volume}{50}}, \bibinfo{pages}{261} (\bibinfo{year}{1978}).

\bibitem[{\citenamefont{Nakano and Toki}(1997)}]{Nakano:1997qd}
\bibinfo{author}{\bibfnamefont{T.}~\bibnamefont{Nakano}} \bibnamefont{and}
  \bibinfo{author}{\bibfnamefont{H.}~\bibnamefont{Toki}}, \bibinfo{journal}{in
  proceedings of International Workshop on Exciting Physics with New
  Accelerators Facilities, SPring-8, Hyogo (World Scientific)}
  p.~\bibinfo{pages}{48} (\bibinfo{year}{1997}).

\bibitem[{\citenamefont{Titov et~al.}(1999)\citenamefont{Titov, Lee, Toki, and
  Streltsova}}]{Titov:1999eu}
\bibinfo{author}{\bibfnamefont{A.~I.} \bibnamefont{Titov}},
  \bibinfo{author}{\bibfnamefont{T.~S.~H.} \bibnamefont{Lee}},
  \bibinfo{author}{\bibfnamefont{H.}~\bibnamefont{Toki}}, \bibnamefont{and}
  \bibinfo{author}{\bibfnamefont{O.}~\bibnamefont{Streltsova}},
  \bibinfo{journal}{Phys. Rev. C} \textbf{\bibinfo{volume}{60}},
  \bibinfo{pages}{035205} (\bibinfo{year}{1999}).

\bibitem[{\citenamefont{Kisslinger and Ma}(2000)}]{Kisslinger:1999jk}
\bibinfo{author}{\bibfnamefont{L.~S.} \bibnamefont{Kisslinger}}
  \bibnamefont{and} \bibinfo{author}{\bibfnamefont{W.-H.} \bibnamefont{Ma}},
  \bibinfo{journal}{Phys. Lett. B} \textbf{\bibinfo{volume}{485}},
  \bibinfo{pages}{367} (\bibinfo{year}{2000}).

\bibitem[{\citenamefont{Llanes-Estrada
  et~al.}(2002)\citenamefont{Llanes-Estrada, Cotanch, de~A.~Bicudo, Ribeiro,
  and Szczepaniak}}]{Llanes-Estrada:2000jw}
\bibinfo{author}{\bibfnamefont{F.~J.} \bibnamefont{Llanes-Estrada}},
  \bibinfo{author}{\bibfnamefont{S.~R.} \bibnamefont{Cotanch}},
  \bibinfo{author}{\bibfnamefont{P.~J.} \bibnamefont{de~A.~Bicudo}},
  \bibinfo{author}{\bibfnamefont{J.~E. F.~T.} \bibnamefont{Ribeiro}},
  \bibnamefont{and} \bibinfo{author}{\bibfnamefont{A.~P.}
  \bibnamefont{Szczepaniak}}, \bibinfo{journal}{Nucl. Phys.}
  \textbf{\bibinfo{volume}{A 710}}, \bibinfo{pages}{45} (\bibinfo{year}{2002}).

\bibitem[{\citenamefont{Ballam et~al.}(1973)}]{Ballam:1973eq}
\bibinfo{author}{\bibfnamefont{J.}~\bibnamefont{Ballam}} \bibnamefont{{\it et~al.}},
  \bibinfo{journal}{Phys. Rev. D} \textbf{\bibinfo{volume}{7}},
  \bibinfo{pages}{3150} (\bibinfo{year}{1973}).

\bibitem[{\citenamefont{Besch et~al.}(1974)}]{Besch:1974rp}
\bibinfo{author}{\bibfnamefont{H.~J.} \bibnamefont{Besch}}
  \bibnamefont{{\it et~al.}}, \bibinfo{journal}{Nucl. Phys.}
  \textbf{\bibinfo{volume}{B 70}}, \bibinfo{pages}{257} (\bibinfo{year}{1974}).

\bibitem[{\citenamefont{Behrend et~al.}(1978)}]{Behrend:1978ik}
\bibinfo{author}{\bibfnamefont{H.~J.} \bibnamefont{Behrend}}
  \bibnamefont{{\it et~al.}}, \bibinfo{journal}{Nucl. Phys.}
  \textbf{\bibinfo{volume}{B 144}}, \bibinfo{pages}{22} (\bibinfo{year}{1978}).

\bibitem[{\citenamefont{Barber et~al.}(1982)}]{Barber:1982fj}
\bibinfo{author}{\bibfnamefont{D.~P.} \bibnamefont{Barber}}
  \bibnamefont{{\it et~al.}}, \bibinfo{journal}{Zeit. Phys. C}
  \textbf{\bibinfo{volume}{12}}, \bibinfo{pages}{1} (\bibinfo{year}{1982}).

\bibitem[{\citenamefont{Anciant et~al.}(2000)}]{Anciant:2000az}
\bibinfo{author}{\bibfnamefont{E.}~\bibnamefont{Anciant}} \bibnamefont{{\it et~al.}},
  \bibinfo{journal}{Phys. Rev. Lett.} \textbf{\bibinfo{volume}{85}},
  \bibinfo{pages}{4682} (\bibinfo{year}{2000}).

\bibitem[{\citenamefont{Barth et~al.}(2003)}]{Barth:2003bq}
\bibinfo{author}{\bibfnamefont{J.}~\bibnamefont{Barth}} \bibnamefont{{\it et~al.}},
  \bibinfo{journal}{Eur. Phys. J. A} \textbf{\bibinfo{volume}{17}},
  \bibinfo{pages}{269} (\bibinfo{year}{2003}).

\bibitem[{\citenamefont{Oh and Bhang}(2001)}]{Oh:2001bq}
\bibinfo{author}{\bibfnamefont{Y.-S.} \bibnamefont{Oh}} \bibnamefont{and}
  \bibinfo{author}{\bibfnamefont{H.~C.} \bibnamefont{Bhang}},
  \bibinfo{journal}{Phys. Rev. C} \textbf{\bibinfo{volume}{64}},
  \bibinfo{pages}{055207} (\bibinfo{year}{2001}).

\bibitem[{\citenamefont{Zhao et~al.}(2001)\citenamefont{Zhao, Saghai, and
  Al-Khalili}}]{Zhao:2001ue}
\bibinfo{author}{\bibfnamefont{Q.}~\bibnamefont{Zhao}},
  \bibinfo{author}{\bibfnamefont{B.}~\bibnamefont{Saghai}}, \bibnamefont{and}
  \bibinfo{author}{\bibfnamefont{J.~S.} \bibnamefont{Al-Khalili}},
  \bibinfo{journal}{Phys. Lett. B} \textbf{\bibinfo{volume}{509}},
  \bibinfo{pages}{231} (\bibinfo{year}{2001}).

\bibitem[{\citenamefont{Titov and Lee}(2003)}]{Titov:2003bk}
\bibinfo{author}{\bibfnamefont{A.~I.} \bibnamefont{Titov}} \bibnamefont{and}
  \bibinfo{author}{\bibfnamefont{T.-S.~H.} \bibnamefont{Lee}},
  \bibinfo{journal}{Phys. Rev. C} \textbf{\bibinfo{volume}{67}},
  \bibinfo{pages}{065205} (\bibinfo{year}{2003}).

\bibitem[{\citenamefont{Williams}(1998)}]{Williams:1998ge}
\bibinfo{author}{\bibfnamefont{R.~A.} \bibnamefont{Williams}},
  \bibinfo{journal}{Phys. Rev. C} \textbf{\bibinfo{volume}{57}},
  \bibinfo{pages}{223} (\bibinfo{year}{1998}).

\bibitem[{\citenamefont{Laget}(2000)}]{Laget:2000gj}
\bibinfo{author}{\bibfnamefont{J.~M.} \bibnamefont{Laget}},
  \bibinfo{journal}{Phys. Lett. B} \textbf{\bibinfo{volume}{489}},
  \bibinfo{pages}{313} (\bibinfo{year}{2000}).

\bibitem[{\citenamefont{Schilling et~al.}(1970)\citenamefont{Schilling,
  Seyboth, and Wolf}}]{Schilling:1970um}
\bibinfo{author}{\bibfnamefont{K.}~\bibnamefont{Schilling}},
  \bibinfo{author}{\bibfnamefont{P.}~\bibnamefont{Seyboth}}, \bibnamefont{and}
  \bibinfo{author}{\bibfnamefont{G.~E.} \bibnamefont{Wolf}},
  \bibinfo{journal}{Nucl. Phys.} \textbf{\bibinfo{volume}{B 15}},
  \bibinfo{pages}{397} (\bibinfo{year}{1970}).

\bibitem[{\citenamefont{Atkinson et~al.}(1985)}]{Atkinson:1984cs}
\bibinfo{author}{\bibfnamefont{M.}~\bibnamefont{Atkinson}}
  \bibnamefont{{\it et~al.}}, \bibinfo{journal}{Z. Phys.}
  \textbf{\bibinfo{volume}{C27}}, \bibinfo{pages}{233} (\bibinfo{year}{1985}).

\bibitem[{\citenamefont{McCormick et~al.}(2004)}]{McCormick:2003bb}
\bibinfo{author}{\bibfnamefont{K.}~\bibnamefont{McCormick}}
  \bibnamefont{{\it et~al.}}, \bibinfo{journal}{Phys. Rev. C}
  \textbf{\bibinfo{volume}{69}}, \bibinfo{pages}{032203}
  (\bibinfo{year}{2004}).

\bibitem[{\citenamefont{Nakano et~al.}(2001)}]{Nakano:2001xp}
\bibinfo{author}{\bibfnamefont{T.}~\bibnamefont{Nakano}} \bibnamefont{{\it et~al.}},
  \bibinfo{journal}{Nucl. Phys.} \textbf{\bibinfo{volume}{A 684}},
  \bibinfo{pages}{71} (\bibinfo{year}{2001}).

\bibitem[{\citenamefont{Ishikawa et~al.}(2005)}]{Ahn:2004id}
\bibinfo{author}{\bibfnamefont{T.}~\bibnamefont{Ishikawa}}
  \bibnamefont{{\it et~al.}}, \bibinfo{journal}{Phys. Lett. B}
  \textbf{\bibinfo{volume}{608}}, \bibinfo{pages}{215} (\bibinfo{year}{2005}).

\bibitem[{\citenamefont{Brun et~al.}(1993)}]{Brun:1978fy}
\bibinfo{author}{\bibfnamefont{R.}~\bibnamefont{Brun}} \bibnamefont{{\it et~al.}},
  \emph{\bibinfo{title}{{CERN} Program Library Long Writeup W5013}},
  \bibinfo{organization}{{CERN} Applications Software Group}
  (\bibinfo{year}{1993}).

\end{thebibliography}

\end{document}